\documentstyle{article}
\begin{document}
\LARGE
\begin{center}
\bf Quantum Creation of BTZ Black Hole
\vspace*{0.6in}
\normalsize \large \rm

Zhong Chao Wu

Dept. of Physics, Beijing Normal University

Beijing 100875, China

and

Dept. of Applied Mathematics, University of Cape Town 

Rondebosch 7700, South Africa

\vspace*{0.4in}
\large
\bf
Abstract
\end{center}
\vspace*{.1in}
\rm
\normalsize
\vspace*{0.1in}

The constrained instanton method is used to study quantum creation of
a $BTZ$ black hole. It is found that the relative creation probability
is the exponential of the negative sum of the entropy associated with the outer and inner black hole horizons. The quantum creations of the 4- or higher dimensional versions of the $BTZ$ black hole
are also studied.

\vspace*{0.3in}

PACS number(s): 98.80.Hw, 98.80.Bp, 04.60.Kz, 04.70.Dy


\vspace*{0.3in}

e-mail: wu@axp3g9.icra.it

\pagebreak

\vspace*{0.3in}

Recently, black holes in lower-dimensional spacetime, especially, the
$BTZ$ black holes in three-dimensional spacetime have received extensive attention [1]. It is hoped that the key features of a higher-dimensional 
black hole could be explored without the unnecessary complications. The main reason is that there does not exist Weyl component in the curvature tensor for lower-dimensional spacetime. If there is no matter field except for a negative cosmological constant, then the spacetime should be locally described as a symmetric space of constant curvature. In fact, it has been shown that the $BTZ$ black hole found under this circumference is the quotient of Anti-de Sitter space under a discrete subgroup of $SO(2,2)$ [2].

In this paper we shall study quantum creation of $BTZ$ black hole in the no-boundary universe [3]. It was thought that, at the $WKB$ level, the seed for a quantum creation of the universe must be an instanton. Later, it was realised that this can  only be applied to the case  with a stationary creation probability. For a general case, the seed must be a constrained instanton [4][5]. The action of a constrained instanton remains stationary only under some constraints. The usual prescription is to find  complex solutions to the Einstein equation and other field equations. Complex analysis is an useful tool in quantum gravity. To respect the principle of general covariance in gravitation theory, one cannot simply work in the complex domain of some privileged but not all spacetime coordinates. If some compact section is of stationary action under the constraints, then it is qualified as a constrained instanton. In general, one can set the constraint that a 3-geometry of the created universe is given. In the no-boundary universe, among these instantons, the one with the maximum stationary action is the seed for the creation [3][5].

The relative
creation probability is the exponential of the negative of the real part of the Euclidean action of the instanton $I = I_r + iI_i$
\begin{equation}
p \approx \exp -I_r .
\end{equation}
The relative probability is for any universe described by
a Lorentzian section analytically continued from the instanton. Thus, the mechanism provides the creation scenario, at the $WKB$ level,
for both open and closed universes [6][7]. 

The Euclidean action  is
\begin{equation}
I = - \frac{1}{2 \pi} \int_M (R - 2 \Lambda + L_m) - \frac{1}{\pi}
\int_{\partial M } K,
\end{equation}
where $\Lambda$ is the cosmological constant, $R$ is the scalar curvature of the spacetime $M$, $K$
is the trace of the second form of the boundary $\partial M$, and
$L_m$ is the Lagrangian of the matter content. If the spacetime is open, then a
boundary term should be added at infinity for variation and the conserved quantities associated with the asymptotic group are therefore defined [8]. Here, we use the units in which $c= k= \hbar = 1$ and $G = 1/8$, for convenience.

The Lorentzian metric of the $BTZ$ black hole can be written [1]
\begin{equation}
ds^2 = - \Delta dt^2 + \Delta^{-1} dr^2 + r^2 (\Omega dt - d\phi)^2,
\end{equation}
where
\begin{equation}
\Delta = -M - \Lambda r^2 + \frac{J^2}{4r^2},
\end{equation}
and
\begin{equation}
\Omega = \frac{J}{2r^2}
\end{equation}
with $ \Lambda < 0, -\infty < t < \infty, 0 < r < \infty$, and $0 \le \phi \le 2 \pi $.

The two constants of integration $M$ and $J$ are identified as mass and angular momentum measured at infinity, respectively [2]. The quantity $\Omega(r)$ is the angular velocity. The quantity $\Delta(r)$ can be factorized as
\begin{equation}
\Delta = - \frac{\Lambda}{r^2} (r - r_1)(r - r_2)(r- r_3)(r -r_4),
\end{equation}
where $r_1 = r_+, r_2 = r_-, r_3 = -r_-, r_4 = -r_+$ and
\begin{equation}
r_{\pm} = \left [ \frac{-M}{2 \Lambda} \left [ 1 \pm \left [ 1 + \Lambda \left [ \frac{J}{M} \right ]^2 \right ]^{1/2} \right ] \right ]^{1/2}.
\end{equation}
We call $r_1, r_2$ the outer and inner black hole horizons, and $r_3, r_4$ are the
negative horizons. In order for these horizons to be real, the conditions
$M>0, -\Lambda J^2 \le M^2$ must be met.

In classical gravity, there is a causal singularity at $r = 0$, and continuing past it would introduce closed timelike lines [2]. Therefore, the region $r<0$ must be cut out from the spacetime. However, for calculations performed in quantum gravity, one does not mind this. People are even more radical to use complex spacetime. Whether or not one accepts the concept of imaginary time as a true entity, one can always use it as a technical tool.

We now construct the constrained instanton which will be the seed for the creation of the $BTZ$ black hole. One can make identification along the imaginary time coordinate $\tau = it$ in the $BTZ$ metric (3), and paste the sector $\tau = \pm \Delta \tau/2$  between two negative horizons $r_3, r_4$ to make a compact manifold. The periodic identification
leads $f_3 (f_4)$-fold cover around the horizon $r_3(r_4)$ in the $(r-\tau)$ plane and forms a cone with a deficit angle there. The deficit angle contribution to the action is the degenerate form of the surface term in Eq.(2).

The surface gravities of the horizons are
\[
\kappa_{3} =\Lambda (r_-^2 - r_+^2)r_-^{-1},
\]
\begin{equation}
\kappa_{4} = \Lambda (r_+^2 - r_-^2)r_+^{-1},
\end{equation}
and the imaginary time periods to avoid the singularities at the horizons are
$\beta_i = 2\pi \kappa^{-1}_i$. Therefore, one has $\Delta \tau = |f_3 \beta_3| = |f_4 \beta_4|$.

The volume contribution to the action is
\begin{equation}
I_v = 2\Lambda(r_+^2 - r_-^2)\Delta \tau
\end{equation}
and the contribution to the action due to the two negative horizons is 
\begin{equation}
I_s = - 4\pi r_4 (1- f_4) - 4\pi r_3 (1- f_3) =  4\pi (r_- + r_+).
\end{equation}

One can get the whole Lorentzian manifold through analytic continuation, beginning with the continuation of time coordinate at the equator, which is the joint section $\tau = 0$ and $\tau = - \Delta \tau/2$, i.e, $\tau =  \Delta \tau/2$. However, the 3-geometry at the equator of the configuration for the wave function is specified for a given differential rotation $\delta$ of the horizons. In order to find the wave function and relative creation probability
for a 3-metric with a given angular momentum, or its canonical conjugate, one has to appeal to a Fourier transformation between the canonical conjugates in the Lorentzian regime [4][5]. Up to a proportionality constant, the period of imaginary time , or the inverse of the temperature, the canonical conjugates in quantum mechanics are equivalent to the thermodynamic conjugates. In the Euclidean regime, at the $WKB$ level, the Fourier transformation is reduced to the Legendre transformation of the action by adding the term
\begin{equation}
I_\delta= J\delta = J (\Omega(r_3) - \Omega (r_4))\Delta \tau
= \frac{J^2 \Delta \tau}{2} \left ( (r_3)^{-2} - (r_4)^{-2} \right )
= 2\Lambda ( r_-^2 - r_+^2) \Delta \tau.
\end{equation}

One obtains the total action from (9)(10)(11)
\begin{equation}
I =  4\pi (r_- + r_+),
\end{equation}
which is independent of the parameter $\Delta \tau$, the only degree left after the 3-metric at the equator is given. Therefore, the manifold pasted is qualified as a constrained instanton. 

One can also use other horizons to construct the constrained instanton. For all
these cases with $r_i, r_j$, the linear terms of $\Delta \tau$ are always cancelled, and the total action is $I = - 4 \pi (r_i + r_j)$. It is the negative of the sum of the entropy of the associated horizons. It is noted that the entropy associated with a horizon is twice the perimeter of the horizon. If one uses the
standard Planckian unit, it becomes one quarter of the perimeter, or ``area'' of the horizon, as usual.

Apparently, one has to use the constrained instanton with $r_3, r_4$ as the seed. At the $WKB$ level, the relative creation probability is the exponential of the negative of the sum of the entropy associated with the black hole outer and inner horizons.  Then the probability is an exponentially decreasing function of the mass parameter $M$. One can get the creation probability for the two extreme cases by letting $M \rightarrow 0$ and $-\Lambda J^2 \rightarrow M^2$.

In gravitational thermodynamics, the Euclidean action is identified as the partition function $Z$ [9]. The entropy $S$ can be evaluated as
\begin{equation}
S = - \frac{\beta \partial}{\partial \beta} \ln Z + \ln Z,
\end{equation}
where $\beta$ is the time identification period.
Thus, the independence of the action from $\beta$ implies that, at the $WKB$ level, the action is the negative of the entropy associated with the horizons for the instanton construction [10]. It seems that the quantum state of the universe is an eigenstate of the entropy operator in the enlarged density matrix representation. This argument can also be applied to the other black hole cases.

The $BTZ$ black hole can be considered as a baby version of the 
Kerr-anti-de Sitter black hole. From the constrained instanton approach [4][5]
, one can find, that the relative probability for Kerr-Newman-(anti-)de Sitter black hole is the exponential of (the negative of) one quarter of the sum of outer and inner black hole horizon areas. The derivation for the Kerr-Newman-anti-de Sitter case is based on the observation that the sum of all horizon areas is constant. The $BTZ$ black hole inherits the similar attribute that the sum of all horizon perimeters is zero.

The 3-dimensional Anti-de Sitter spacetime can be considered as a ``bound state'', its metric can be written [1]
\begin{equation}
ds^2 = -(1 - \Lambda r^2) dt^2 + (1 - \Lambda r^2)^{-1} dr^2
+ r^2 d \phi^2.
\end{equation}
It looks like that the bound state has $M= - 1, J = 0$. It is separated from the black hole spectrum by a mass gap. One can use the two imaginary horizons $r = \pm \Lambda^{-1/2}$ to construct the constrained instanton, the action is 0. Or one can use one of the horizons and the origin $r = 0$ to make a constrained instanton, the action is imaginary. Therefore, the relative creation probability, at the
$WKB$ level, is 1.

It may also be appropriate to study the 4-dimensional version of the $BTZ$ solution. It is the constant curvature black hole with metric
\begin{equation}
ds^2 = -\frac{3}{\Lambda} \Delta (- \sin^2 \theta dt^2 + d\theta^2) - \frac{3}{\Lambda} \Delta^{-1} dr^2 + r^2 d \phi^2,
\end{equation}
where $ 0\le \phi \le 2 \pi, 0 \le \theta \le \pi$ and
\begin{equation}
\Delta = (r^2 - r_0^2) r_0^{-2}.
\end{equation}

This spacetime can be considered as the quotient of Anti-de Sitter space under a discrete subgroup of $SO(3,2)$. Closed timelike curves are introduced by the identification. The horizon $r = \pm r_0$ are the chronological singularities. The horizon is the product of a null conoid in the $(t-\theta - r)$ space and a circle in the $\phi$ space. We switch back to the standard Planckian unit here.

One can set $\tau = it$, and identify the imaginary time $\tau$ with period $\Delta \tau = 2\pi $ between one of the two horizons $ r = \pm r_0$ and the origin $r = 0$. This makes the $(\tau - \theta - r)$ section a regular 3-sphere. We also set $\chi = i \phi$, and identify $\chi$ with period $2 \pi |\Lambda |^{-1/2} r^{-1}_0$. The period of the $\chi$ coordinate is determined by the regularity at $r =0$. This constructed manifold is the standard $S^4$ space with signature $(----)$. With respect to the signature of the metric, the effective cosmological constant becomes positive. The action is a constant $-3\pi/\Lambda$. 
However, the $S^4$ instanton is not of the largest real part of the constrained instanton action. If we repeat the process for the $t$-coordinate and leave the $\phi$-coordinate intact, we obtain a constrained instanton with signature $(---+)$ and the action is purely imaginary. This is the creation seed and the relative probability is 1. The creation of a 4-dimensional $BTZ$ black hole with a domain wall for compactification has been studied [12].

The origin of this phenomenon stems from the fact for the $n$-dimensional $BTZ$ black hole, the resulting topology of the identification is $R^{n-1} \times S^1$, the signature of the spacetime changes when crossing the horizon for $n \ge 4$. There does not exist a constrained instanton similar to that responsible for the creation in the $n=3$ case. For $n \ge 3$, one can obtain the instanton which is the product of the $(n-1)$-sphere $S^{n-1}$ with signature $(\overbrace{-\cdots -}^{n-1})$ and the circle $S^1$ with signature $(+)$ as shown in the case $n= 4$ above. One leaves the coordinate $\phi$ at $S^1$ intact. For even $n$ one gets the imaginary action, the probability is 1; for odd $n$, the action is negative and the calculation of the creation probability is straightforward.

\vspace*{0.4in}

\bf Acknowledgements
\rm
\vspace*{0.1in} 

I would like to thank G.F.R. Ellis of University of Cape Town for his hospitality.
\vspace*{0.1in}
  
\bf References:

\vspace*{0.1in} 
\rm

1. M. Ba$\tilde{n}$\rm ados, C. Teitelboim and J. Zanelli, 
\it Phys. Rev. Lett. \bf \rm \underline{69}, 1849 (1992).

2. M. Ba$\tilde{n}$\rm ados, M. Henneaux, C. Teitelboim and J. Zanelli, 
\it Phys. Rev. \bf D\rm \underline{48}, 1506 (1993).

3. J.B. Hartle and S.W. Hawking, \it Phys. Rev. \rm \bf D\rm
\underline{28}, 2960 (1983).

4. Z.C. Wu, \it Int. J. Mod. Phys. \rm \bf D\rm\underline{6}, 199
(1997), gr-qc/9801020.

5. Z.C. Wu, \it Phys. Lett. \bf B\rm 
\underline{445}, 274 (1999); gr-qc/9810077. 

6. Z.C. Wu, \it Phys. Rev. \bf D\rm 
\underline{31}, 3079 (1985). 

7. S.W. Hawking and N. Turok, \it Phys. Lett. \bf B\rm 
\underline{425}, 25 (1998), hep-th/9802030.

8. T. Regge and C. Teitelboim, \it Ann. Phys. \rm (N.Y.) \rm \underline{88}, 286 (1974).

9. G.W. Gibbons and S.W. Hawking, \it Phys. Rev. \bf D\rm 
\underline{15}, 2725 (1977).

10. Z.C. Wu, \it Gene. Relativ. Grav. \rm \underline{31}, 1097 (1999), gr-qc/9812051.

11. M. Ba$\tilde{n}$\rm ados, \it Phys. Rev. \bf D\rm 
\underline{57}, 1068 (1998).

12. R.B. Mann, \it Nucl. Phys. \bf B \rm \underline{516}, 357 (1998).

\end{document}